# Updating the Near-Earth Neutral Line Model

*Wolfgang Baumjohann & Rumi Nakamura*

*Institut für Weltraumphysik der Österreichischen Akademie der Wissenschaften*

*Schmiedlstrasse 6, A-8042 Graz, Austria*

Email: baumjohann@oeaw.ac.at, rumi@oeaw.ac.at

**Abstract**

In the past thirty years the Near-Earth Neutral Line model for substorms has been proven to be correct in its basic assumption, i.e., that a new neutral line is created near the Earth during substorm onset. Yet there have also been numerous changes and additions to this model, which were necessary to give a more complete and deeper understanding of the substorm process. This process is still continuing.

## 1. Introduction

More than three decades ago, the first phenomenological description of substorm in terms of auroral activity was introduced (Akasofu, 1964). The auroral substorm concept was extended soon after to encompass disturbances throughout the entire magnetosphere-ionosphere system. The energy for substorms was attributed to a physical process known as magnetic reconnection, envisioned to convert energy between magnetic field and charged particles by cutting magnetic field lines and rejoining them with different ones.

Thirty years ago, McPherron et al. (1970) and Hones et al. (1970) have put together the first aspects of a phenomenological substorm model, which had as key ingredient the development of a new magnetic neutral line closer to the Earth (than the distant neutral line at about 80-100 $R_E$) during substorm onset, and was subsequently called Near-Earth Neutral Line model, or short NENL model. This model has now been around for more than 25 years. During these years, it has been evolved considerably. While it has been thought earlier that the new neutral line is located rather close to the Earth, results from the recent Geotail mission have now established that it is typically located between 20 and 25 $R_E$, since it is this distance range where the high-speed plasma flow typically changes from Earthward to tailward during substorm onset (see Fig. 1 and, e.g., Nagai et al., 1998; Baumjohann et al., 1999).

Actually most substorm researchers, even those with competing models like Rostoker (1996) or Lui and Murphree (1998), do agree that such a neutral line appears during a substorm. Near-Earth reconnection is essential for shortening the magnetic flux tube, i.e., the bundle of field lines filled with collisionless plasma and thus tied closely together, before it can proceed to complete its Earthward convection cycle from the high-β magnetotail into the inner dipolar magnetosphere.

In the present short review on the present state of the NENL model, I will not much go into the history of its development over the past 25-30 years, but rather focus on new elements and additional insight gained during the past ten years. Actually, the last 2-3 years were the most interesting ones for the development of this model, with many important new facets of this model being explored, due to, on the one hand, much improved data, mainly from the Japanese Geotail mission (Nishida, 1994), and, on the other side, more sophisticated numerical simulations of the processes in the Earth's magnetotail (e.g., Birn et al., 1999).

As most reviews, also this one will be biased to the author's own views and work and I would like to apologize to all those colleagues, may they be NENL model proponents or opponents, whose contributions are not acknowledged here to the extent they should. For other substorm models, I refer the reader to Rostoker (1996) or Lui and Murphree (1998) and references therein. A slightly different view on the present state of the NENL model can be found in, for example, Baker et al. (1996, 1999).

## 2. Substorms as a natural way to avoid the pressure crisis

During a magnetic substorm, enhanced solar wind energy input and dayside magnetopause reconnection lead to magnetic field lines being transported over the poles into the tail lobes. This is the growth phase, which can easily be observed during isolated substorms. These field lines, or part of thereof, are subsequently reconnected at the distant neutral line. From here, the field lines and the plasma tied to them start traveling Earthward with high velocities.

Since the magnetotail tailward of, say, 30 $R_E$ is a high-$\beta$ region, where the plasma energy clearly dominates over the magnetic energy, this flow proceeds rather uninhibited (see Fig. 2). However, when it reaches about 25 $R_E$, plasma-$\beta$ starts to drop and the influence of the Earth's dipolar field increases. For the flow to proceed adiabatically further inward, the plasma pressure inside a flux tube would have to increase dramatically, since the flux tube volume strongly decreases close to the Earth.

In fact, the flux tube would be repelled from the inner magnetosphere by its own over pressure. In order to avoid this pressure catastrophe, near-Earth reconnection sets in, cuts the field lines and thus severely reduces the flux tube volume, and thus its mass and particle content, further (see Fig. 3 and, e.g., Hesse et al., 2001). This is the start of the expansion phase. The outer part of the flux tube moves outward again as a bundle of closed field lines, the plasmoid, while the inner part proceeds in its convection cycle.

## 3. Transient nature of the flux transport

While it has earlier been thought that this NENL is always a large-scale phenomenon, the intermittent, bursty nature of the high-speed flow between 10 and 20 $R_E$ strongly suggests that near-Earth reconnection also occurs in an intermittent and bursty fashion and that it may often be a localized phenomenon.

In fact, Baumjohann et al. (1990) showed that in the neighborhood of the neutral sheet inside of 20 $R_E$ high-speed flows are rather bursty with the majority staying uninteruptedly at high speed levels for no more than 10 sec. Angelopoulos et al (1992) noted that the high-speed flows did not occur at random, but rather organize themselves in 10-min time scale flow enhancements which they called bursty bulk flow events. Fig. 4 shows a superposed epoch analysis of the behavior of the flows speed and the magnetic field elevation around the center of more than 100 bursty bulk flow events. The typical bursty flow event lasts about 10 min and is often associated with a dipolarization of the ambient magnetic field, a feature which will become important in Section 5.

Despite the fact that the bursty flows typically cover only 10-20% of all measurements, they are the primary means of Earthward transport of mass, momentum, and magnetic flux (e.g., Angelopoulos et al., 1999; Schödel et al., 2000). The relative numbers vary somewhat with radial distance, but in the midnight sector around 20 $R_E$, where the bursty transport maximizes, it often accounts for more than 80% of the Earthward transport of mass, momentum, and magnetic flux.

While bursty flow events are always seen during substorm onset and the expansion phase (if a satellite is in the right location to observe them), it has always been argued that many of them occur without classical substorm signatures. However, those signatures may have simply been overlooked. Taking into account that even pseudo-breakups and very small substorms show all the signatures of near-Earth reconnection (see, e.g., Aikio et al., 1999; Petrukovich et al., 1998, 2000) and should thus be classified as substorms, and that there is a one-to-one correlation between auroral brightening and fast flows (Nakamura et al., 2000), it seems likely that there is a one-to-one correlation between bursty flow events and substorm activity.

## 4. Substorms with closed flux reconnection only

While in the original classical version of the NENL model the energy dissipated during the substorm expansion phase was thought to come directly from the magnetic energy stored in the tail lobes via reconnection of these open field lines, during recent years it has it has been noticed that near-Earth reconnection is often restricted to the plasma sheet, i.e. closed field lines, and only in major substorms lobe field lines are reconnected at the near-Earth neutral line (e.g., Baumjohann et al., 1996; Fox et al., 1999). Hence, reconnection of open flux stored in the tail lobes can not be considered anymore as a necessary element of a substorm, at least not for all of those that occur during disturbed intervals where often one substorm follows the other.

Of course, the energy dissipated during a substorm still mainly comes from magnetic energy stored as open flux in the tail lobes during the growth phase. But during a lot of substorms, and perhaps even during the first phase of those with clear open flux reconnection signatures, magnetic-to-plasma energy conversion will take place at the distant neutral line, while near-Earth reconnection will mainly be driven by the enhanced transport in the distant tail and the need to complete the convection cycle yet to avoid the impeding pressure crisis (Hesse et al., 2001).

If substorms come in a sequence, as they often do, the initial pseudo-breakups (see, e.g., Nakamura et al., 1994; Aikio et al., 1999) and the initial full substorm onsets show signatures of reconnection, but do not seem to be associated with open flux reconnection (e.g., Fox et al., 1999; Mishin et al., 2000). Open flux reconnection rather seems to occur for the last substorm in that sequence, which starts when the interplanetary magnetic field finally turns northward (Caan et al., 1975; Mishin et al., 2000).

## 5. Flow braking and ionospheric substorm signatures

The fast bursty flows generated by near-Earth reconnection soon encounter another obstacle: the strongly dipolar field of the inner magnetosphere, where plasma-β drops below unity (see Fig. 2). The outward pressure gradient strongly decelerates and brakes this flow (Shiokawa et al., 1997), with possibly some of it diverted duskward. As sketched in Fig. 5, the braking itself, and even more the plasma pressure gradients built up by the plasma transported inward with the bursty flows, leads to the generation of a dawnward cross-tail current, diversion of the originally duskward neutral sheet current through the auroral ionosphere, dipolarization of the field (noticeable in Fig. 4), and thus the establishment of the substorm current wedge (Birn et al., 1999). The buildup of the substorm current wedge will be associated with Pi2 pulsations and auroral kilometric radiation (Shiokawa et al., 1998; Fairfield et al., 1999).

Furthermore, there are ideas that the stress put onto the field lines during the braking and/or diversion will lead to parallel electric fields and thus brightening of aurora (Haerendel, 1992). Brightening of aurora associated with braking of bursty fast flows has actually been observed recently (Fairfield et al., 1999, Nakamura et al., 2000). Hence, the flow braking may cause all the typical signatures of substorm onset in the ionosphere, at latitudes corresponding to L-values around 10.

The bursty bulk flow and its braking is one of the key ingredients of the present-day near-Earth neutral line model (depicted in Fig. 5) and distinguishes it from the classical one which had the auroral breakup and substorm current wedge tied to a near-Earth X-line. It also resolves a long-standing puzzle (and objection against the original classical version), namely that the field lines where near-Earth reconnection occurs should have their footprint near the breakup aurora. This was difficult to reconcile, since the field lines threading the breakup aurora typically map to the equatorial region near 10 $R_E$ rather than to 20-25 $R_E$. Now we can say that near-Earth reconnection takes place further out and that the brightening of aurora is related to the fast bursty flows and occurs somewhere along their pass, most likely where the fast flows generated by near-Earth reconnection are effectively stopped and perhaps partially diverted. However, it remains unclear how this location is connected to the most equatorward auroral arc, where the auroral breakup typically occurs.

## 6. Substorm Recovery and quenching of near-Earth reconnection

During the course of the substorm, the fast bursty flows will transport more and more magnetic flux inward. This flux is being piled-up and the associated dipolarization front moves tailward. As shown in Fig. 6, between 30-45 min after onset the tailward expanding dipolarization front reaches the location of the near-Earth neutral line near 20-25 $R_E$. Since reconnection cannot proceed in a dipolar magnetic field configuration, the near-Earth neutral line moves rapidly tailward, still producing fast Earthward flow. This is the beginning of the recovery phase (Baumjohann et al., 1999).

Hence, near-Earth reconnection effectively strangles itself after some time. It is quenched by the successive pile-up of those closed dipolar field lines it generates. This is the simple solution of the long-standing puzzle why often the unloading of stored magnetic energy stops before the energy reservoir available in the magnetic tail lobes is fully depleted. It may also answer the question posed in Section 5, i.e., why near-Earth reconnection during many substorms just reconnects closed plasma sheet field lines and does not always proceed to reconnect open lobe field lines at the near-Earth neutral line: at least in these cases the dipolar flux pile-up reaches the near-Earth neutral line earlier than reconnection at the latter reaches the lobe.

## 7. Conclusion and outlook

While the present-day version of the Near-Earth Neutral Line model is the most comprehensive description of the processes occurring during a substorm, i.e., fits most aspects of the observed data and agrees with numerical simulations of magnetotail processes, there are still a couple of open questions.

One key question is what determines the onset of near-Earth reconnection, especially why does it occur at 20-25 $R_E$ and why is it so time-dependent and/or localized (as indicated by the bursty nature of the fast flows). Is it really because plasma-$\beta$ starts to drop here as indicated in Fig. 2 and discussed above? Do local conditions play an important role?

Another one is why substorms come in so vastly different sizes, yet even very small ones show the basic signatures of near-Earth reconnection and plasmoid release (see, e.g., Aikio et al., 1999; Petrukovich et al., 2000). Why is reconnection sometimes quenched before it reaches open lobe field lines? Does it always choke itself as speculated in the previous section? How can plasmoids be released under such conditions, when closed field lines may still exist tailward of the plasmoid?

Another important still somewhat open issue pertains to the connection between the magnetotail outside of, say, 10 $R_E$ and the inner magnetosphere and ionosphere. While the flow braking scenario was a first breakthrough in answering this question, it still leaves a couple of open issues. Among them are (1) a possible active role of the ionosphere in something like a feedback mechanism by, for example, the generation of a Cowling channel in the active aurora region (Baumjohann et al., 1981) and (2) a more violent reaction of the inner magnetosphere around 6-8 $R_E$ to the flow braking than presently envisioned in the NENL model (see, for example, Ohtani et al., 1999; Lui et al., 2000).

While some of these questions may be answered in the near future by more advanced numerical modeling (using global hybrid or even kinetic codes with enough spatial resolution instead of the present-day MHD simulations) or more ingenious analysis of presently available data, some may have to wait for new data to arrive from multipoint satellite missions. Unfortunately, the orbit of the first of the four spacecraft missions, ESA's Cluster II, is not well suited for magnetotail physics, but NASA's planned MMS mission will certainly be of great help in studying the microphysics of near-Earth reconnection. However, the more mesoscale aspects of substorm physics, e.g. the turbulent aspects of the bursty fast flows, will have to await truly multipoint missions, like the Draco mission presently under study.

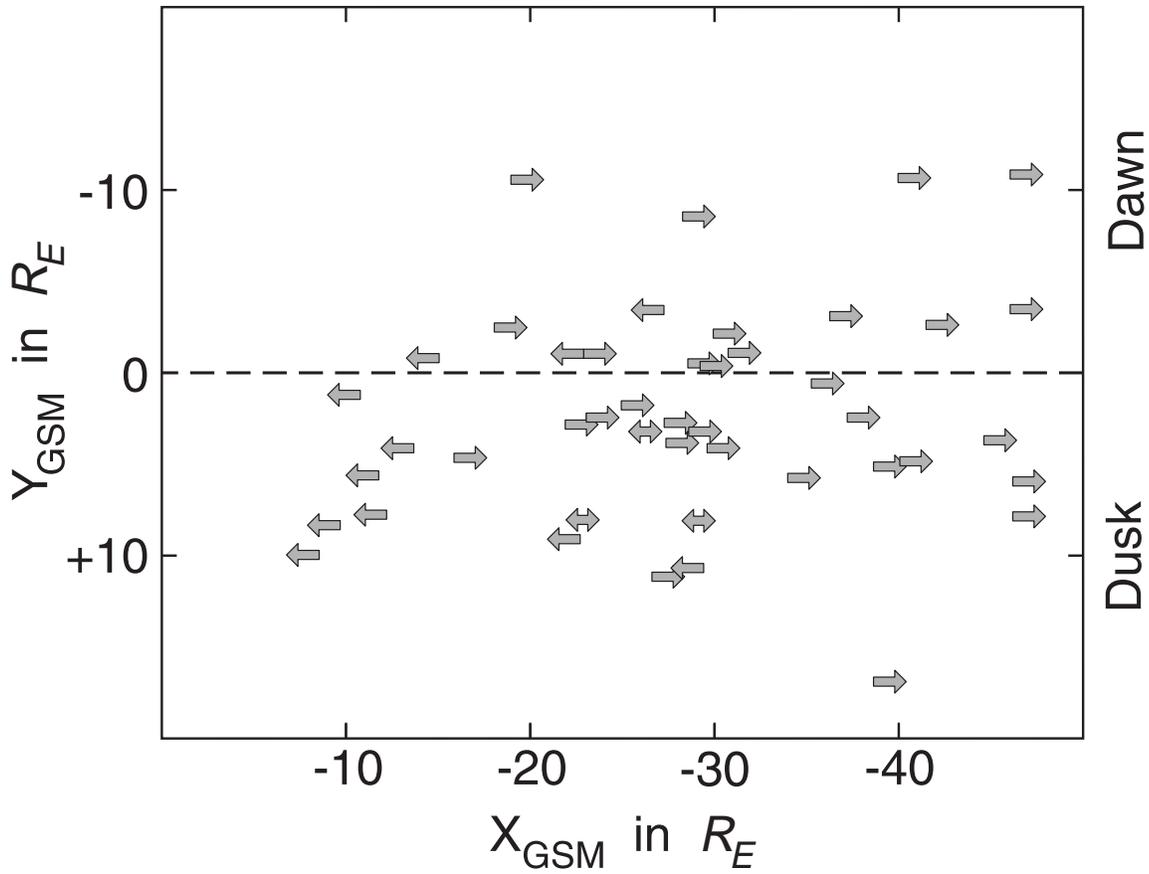

*Fig. 1: Reversal of the dominant direction of fast flows during substorm onset, typically observed by Geotail between 20 and 25 $R_E$ (after Nagai et al., 1998).*

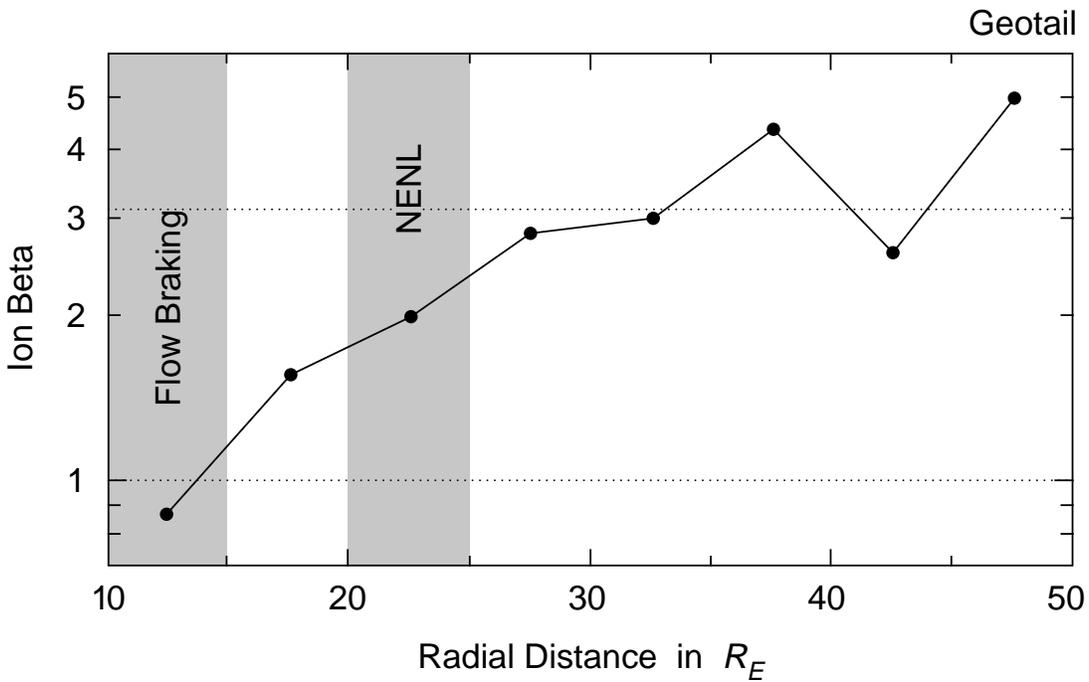

*Fig. 2: Average radial profile of ion beta in the magnetotail based on Geotail data.*

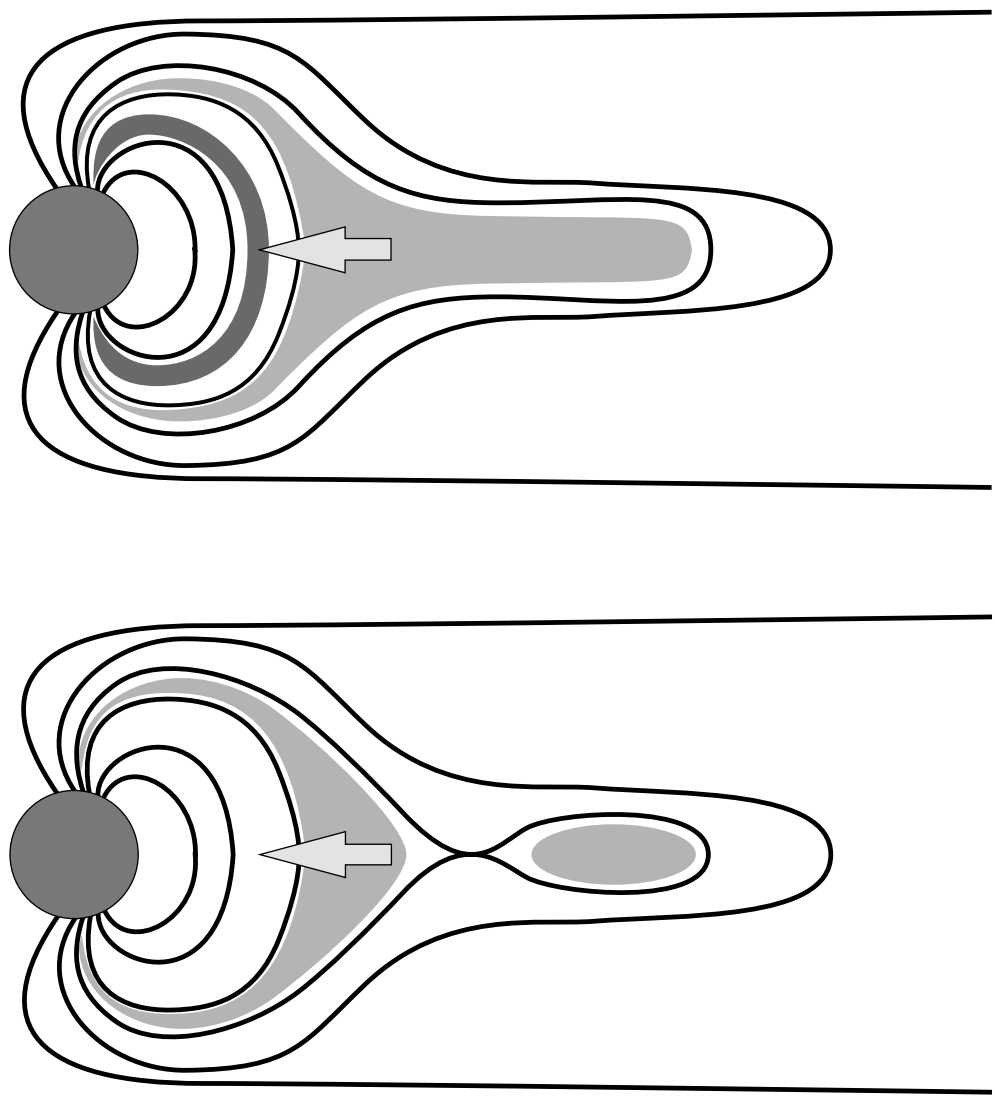

*Fig. 3: Earthward convection and avoidance of a pending pressure catastrophe by plasma sheet thinning and near-Earth reconnection (after Hesse et al., 2001).*

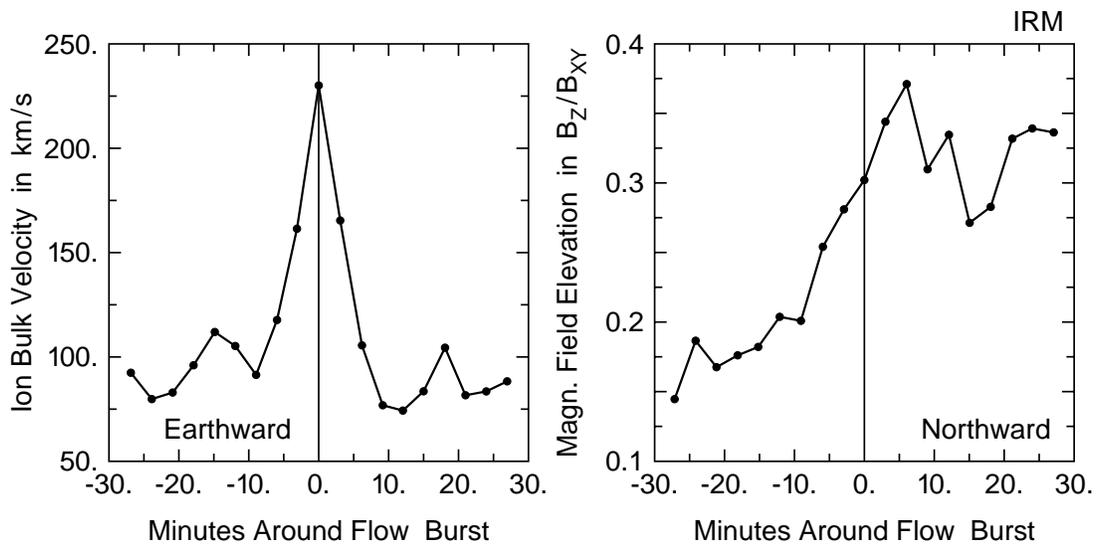

*Fig. 4: Typical variation of Earthward flow velocity and northward magnetic field dipolarization during bursty fast flow events (from Baumjohann, 1993).*

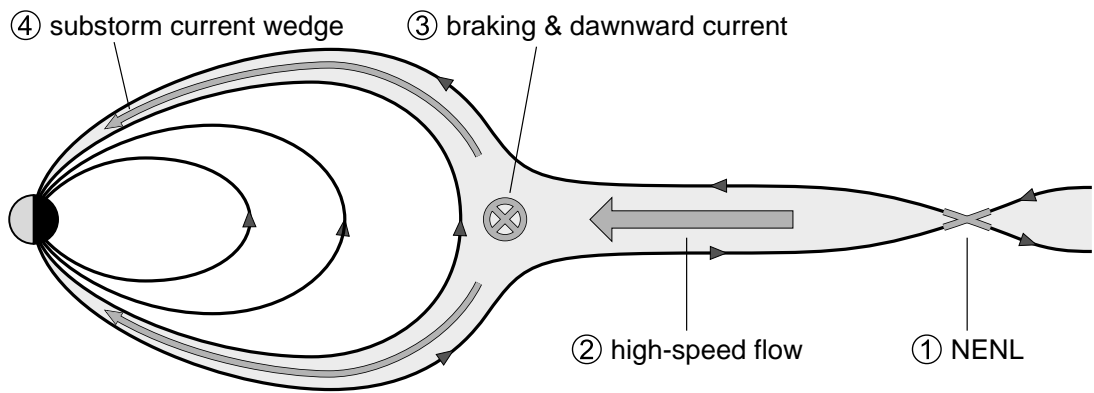

*Fig. 5: Sequence of events during substorm onset (after Shiokawa et al., 1998).*

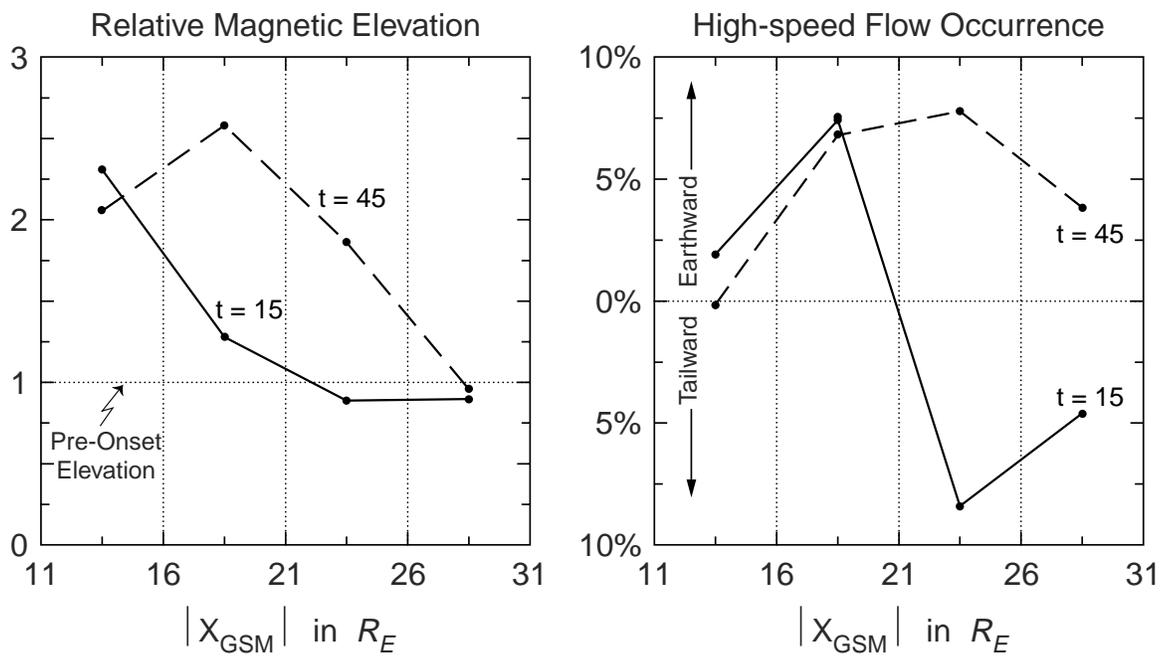

*Fig 6: Radial profiles of (normalized) magnetic field elevation and fast flow occurrence rates during the expansion and the recovery phase of an average substorm (simplified from Baumjohann et al., 1999).*